\documentclass[aps,pra,twocolumn,showpacs,groupedaddress]{revtex4}
\usepackage{amssymb}

\usepackage{graphicx}
\begin{document}
\title{Optimal Lyapunov quantum  control on two-level systems:
convergence and extended techniques}

\author{L. C. Wang}
\email{wanglc@dlut.edu.cn}
\affiliation{School of Physics and Optoelectronic Technology,
              Dalian University of Technology, Dalian 116024, China}

\author{S. C. Hou}
\affiliation{School of Physics and Optoelectronic Technology,
              Dalian University of Technology, Dalian 116024, China}
\author{X. X. Yi}
\email{yixx@dlut.edu.cn} \affiliation{School of Physics and
Optoelectronic Technology,
              Dalian University of Technology, Dalian 116024, China}
\author{Daoyi Dong}
\email{daoyidong@gmail.com} \affiliation{ School of Engineering and
Information Technology, University of New South Wales at the
Australian Defence Force Academy, Canberra, ACT 2600, Australia}

\author{Ian R. Petersen}
\email{i.r.petersen@gmail.com}
\affiliation{ School of  Engineering and Information Technology,
University of New South Wales at the Australian Defence Force Academy,
Canberra, ACT 2600, Australia}

\date{\today}
\begin{abstract}
Taking a two-level system  as an example, we show that a strong control
field may enhance the efficiency of optimal Lyapunov quantum control in [Hou et al., Phys. Rev. A \textbf{86}, 022321 (2012)] but could decrease
its control fidelity. A relationship between the  strength of the
control field and the control fidelity is established. An extended technique, which combines   free evolution
and external control, is proposed to improve the control fidelity. We analytically demonstrate that the extended technique can be used to
design a control law for steering a
two-level system exactly to the target state. In such a way, the convergence of the
extended optimal Lyapunov quantum control can be guaranteed.
\end{abstract}

\pacs{03.67.-a, 02.30.Yy}
\maketitle

\section{Introduction}
Quantum information theory as an
interdisciplinary research field has rapidly grown in the past decades \cite{nielsenbook}. Quantum
control theory, the application of control theory to quantum systems, has
attracted much attention due to its potential applications in
quantum information
theory \cite{alessandro,wisemanbook,dong2010-2,dong2010-1}. The main
goal in quantum control  theory is to establish a  theoretical
footing and develop a series of systematic methods for  active
manipulations and control of quantum systems. Lyapunov quantum control provides a systematic design method for some difficult quantum control
tasks \cite{dong2010-1,altafini2007}.
It uses feedback design to construct control
fields but applies the fields to  a quantum  system in an open-loop
way. It provides us with a simple  way to
design control fields for the manipulation of quantum state transfer
\cite{grivopoulos2003,mirrahimi2005,wangx2010,wangx2009, kuang2008,
wangw2010,yi2011-1,coron2009,beauchard2007,yi2011-2,yi2009,yi2010,hou2012}.

Although  much progress has been made  in research on Lyapunov quantum control, techniques to speed up Lyapunov quantum control have rarely
been presented. Study of this problem is helpful to shorten the
control time and hence to reduce the decoherence effect induced by
environments. Recently, an optimal method has been
proposed to speed up Lyapunov quantum control \cite{hou2012},
where a design approach was presented to make the Lyapunov
function decrease faster. However, the results in \cite{hou2012} were demonstrated only based on
numerical simulations and the
convergence of such an optimal Lyapunov method was not completely analyzed.

In order to clearly show the essence of the optimal Lyapunov quantum
control method and to explore the possibility to  improve this control approach, we
 present an exactly solvable model to study this problem  in
this paper. We observe  that the method of optimal Lyapunov control \cite{hou2012} leads to a
limit on the control fidelity: a stronger control field can enhance the
efficiency of the control method, but it could decrease the control fidelity. The
convergence time of the Lyapunov function is closely related to the strength
of the control field. This fact demonstrates that the convergence is dependent
on the strength of the control field.  Stimulated by these
observations, we propose an extended method of optimal Lyapunov control that
combines free evolution and external control for quantum systems. We show
that the extended technique can guarantee convergence and make the
convergence independent of the strength of the control field.

This paper is organized as follows. In Sec. {\rm II}, we   present a
brief review on  the optimal Lyapunov control method proposed  in
\cite{hou2012}.  In Sec. {\rm III}, we analytically calculate
the control fidelity for a two-level system. A limit to the control
fidelity is presented and   a relationship between the limit and the strength of the
control field is established. An extended
technique is
proposed to guarantee the convergence of optimal Lyapunov control in Sec. {\rm IV}. Conclusions are presented in Sec. {\rm V}.

\section{optimal Lyapunov quantum control}
In Lyapunov quantum control, the system  is steered from an initial
state to a target state by control fields determined by a Lyapunov
function $V$, which should decrease with time and
converge to its minimum. Considering a closed quantum system, its state
$\rho$ evolves as
\begin{eqnarray}
\frac{d\rho}{dt}=-i[H_0+H_C(t),\rho],
\end{eqnarray}
where $H_0$ is the free Hamiltonian of the system, and
\begin{eqnarray}
H_C(t)=\sum_{n=1}^kf_n(t)H_n
\end{eqnarray}
denotes the control Hamiltonian. Assume that $|f_{n}(t)|\leq S$ and $P$ is a  positive Hermitian operator
which carries information of the target state.  The Lyapunov
function  can be  defined as
\begin{eqnarray}
V=\text{Tr}(P\rho).
\end{eqnarray}
The time derivative of  the Lyapunov function is given by (assuming $[H_0,P]=0$)
\begin{eqnarray}
\dot{V}=\text{Tr}(-iP[H_0+\sum_{n=1}^kf_n(t)H_n,\rho])
=\sum_{n=1}^kf_n(t)T_n,
\end{eqnarray}
where $T_n=\text{Tr}(-i\rho [P, H_{n}])$. In order to find the control fields $f_n(t)$ that steer the Lyapunov  function to
its minimum  as fast as possible, the control fields can be selected as follows:
\begin{eqnarray}
f_n(t)=\left\{
\begin{array}{l}
-S,  \ \ \ \  (T_n>0),\\
0,  \ \ \ \ \ \ \ (T_n=0),\\
S, \ \ \ \ \ \ (T_n<0).
\end{array}
\right.
\end{eqnarray}
Substituting the control fields into the time derivative of the
Lyapunov function, we have,
\begin{eqnarray}
\dot{V}=\sum_{n=1}^kf_n(t)T_n=-S\sum_{n=1}^k|T_n|.
\end{eqnarray}
It is clear that  $\dot{V}\leq 0$, which ensures the decreasing of the
Lyapunov function.

In \cite{hou2012}, the method of optimal Lyapunov control has been
applied to a three-level system. Numerical results showed that the system could be steered optimally into the
target state with high fidelity. Observing the numerical results in
\cite{hou2012}, we find that the control fields could change signs
between ``positive" and ``negative" values very frequently
when the system is very close to the target state.  This indicates
that the optimal Lyapunov control with finite strengths of control
fields may  not ensure the convergency near the target state. To
clarify this point, an analytical investigation is necessary. In the next
section, we focus on this issue using an exact model of a two-level system.

\section{optimal Lyapunov control on two-level systems}

Now, we apply the optimal Lyapunov
control method in Sec. II to a  two-level system. A relationship between the strength of
the control field and the control fidelity can be found by analyzing the time evolution of the system under the Lyapunov control.

\subsection{Evolution operator}
Consider a two-level system governed by the following  Hamiltonian
\begin{eqnarray}
H=\frac {\omega} 2\sigma^z+f\sigma^x,
\end{eqnarray}
where we set $\hbar=1$. $\omega$ is the level spacing of the system,
$f=f(t)$ denotes the control field. Assume that the aim is to steer the system
from  an arbitrary state
$|\psi_0\rangle=\cos\frac{\gamma_0}2|e\rangle+e^{i\phi}\sin\frac{\gamma_0}2|g\rangle$
to  state $|e\rangle$ (target state), where $|g\rangle$ is the ground state of the system, $|e\rangle$ is the excited state,
$\gamma_{0}\in [0, \pi]$ and $\phi \in [0, 2\pi]$. Define a positive operator
\begin{eqnarray}
P_g=I-|e\rangle\langle e|=|g\rangle\langle g|.
\end{eqnarray}
The  Lyapunov function can be written as
\begin{eqnarray}
V_g=\text{Tr}[P_g\rho],
\end{eqnarray}
with
\begin{eqnarray}
\rho=|\psi\rangle\langle \psi|,  \ \ \ \ \ \
|\psi\rangle=a(t)|e\rangle+b(t)|g\rangle.
\end{eqnarray}
The Lyapunov function $V_g$ represents the overlapping between  the
function $I-|e\rangle\langle e|$ of target state $|e\rangle$ and the
actual state of the system. The time derivative of the Lyapunov
function can be calculated as follows (with abbreviations, $a=a(t)$,
$b=b(t)$):
\begin{eqnarray} \label{dotV}
\dot{V}_g&=&\text{Tr}[P_{g}\dot{\rho}]=
\text{Tr}\{-iP_{g}[\frac{\omega}2\sigma^z+f\sigma^x,\rho]\}  \nonumber\\
&=&\text{Tr}\{-iP_{g}[\frac{\omega}2\sigma^z,\rho]\}
+\text{Tr}\{-iP_{g}[f\sigma^x,\rho]\} \nonumber\\
&=&2f\cdot\text{Im}(ab^*). \label{dV}
\end{eqnarray}
If $\dot{V}_g\leq 0$  for all times,  $V_g$ would monotonically
decrease with time under the control,  meanwhile the system is
asymptotically steered into the target state $|e\rangle$. Using the
method in Sec. {\rm II}, the control field $f(t)$ takes values
\begin{eqnarray} \label{qubitcondition}
f(t)=\left\{
\begin{array}{l}
S,   \ \ \ \ \ \ \    \text{Im}(ab^*)<0   ,\\
0 ,  \ \ \ \ \ \ \    \text{Im}(ab^*)=0   ,\\
-S , \ \ \ \ \        \text{Im}(ab^*)>0.
\end{array}
\right. \label{con_law}
\end{eqnarray}
It is clear that the control field in (\ref{qubitcondition})
guarantees $\dot{V}_g\leq 0$.

With the optimal Lyapunov control, the time evolution of the two-level system can be analytically
calculated. In a basis spanned by  $\{|e\rangle, |g\rangle\}$, the total
Hamiltonian can be expressed as
\begin{eqnarray}
H=\sqrt{\frac{\omega^2}4+f^2}
\left(  \begin{array}{cc}
\cos\theta & \sin\theta\\
\sin\theta  & -\cos\theta
\end{array} \right),
\end{eqnarray}
with $\theta$ defined by
$$\tan\theta=\frac{2f}{\omega}.$$
The eigenvalues of the Hamiltonian $H$ are
$$E_{\pm}=\pm\sqrt{\frac{\omega^2}4+f^2},$$
and the corresponding eigenvectors are given by,
\begin{eqnarray}
|E_+\rangle&=&\cos\frac{\theta}2|e\rangle +\sin\frac{\theta}2|g\rangle, \nonumber\\
|E_-\rangle&=&\sin\frac{\theta}2|e\rangle-\cos\frac{\theta}2|g\rangle .
\end{eqnarray}
The time evolution operator can be calculated to be
\begin{widetext}
\begin{eqnarray}
U&=&\exp{(-iHt)}  =\left(
\begin{array}{cc}
e^{-iE_+t}\cos^2\frac{\theta}2+e^{-iE_-t}\sin^2\frac{\theta}2
&\frac 1 2(e^{-iE_+t}-e^{-iE_-t})\sin\theta \\
\frac 1 2(e^{-iE_+t}-e^{-iE_-t})\sin\theta
& e^{-iE_-t}\cos^2\frac{\theta}2+e^{-iE_+t}\sin^2\frac{\theta}2
\end{array}
\right).
\end{eqnarray}
\end{widetext}
In the absence of a control field (i.e., $f=0$), we have $\theta=0$.
The time evolution operator reduces to a diagonal form,
\begin{eqnarray}
U=\left(
\begin{array}{cc}
e^{-i\omega t/2}  & 0 \\
0   & e^{i\omega t/2}
\end{array}
\right).
\end{eqnarray}
In the following, we use the evolution operator $U$ to calculate
the state of the system under control.

\begin{figure}
\includegraphics[width=0.95\columnwidth,height=0.8\columnwidth]{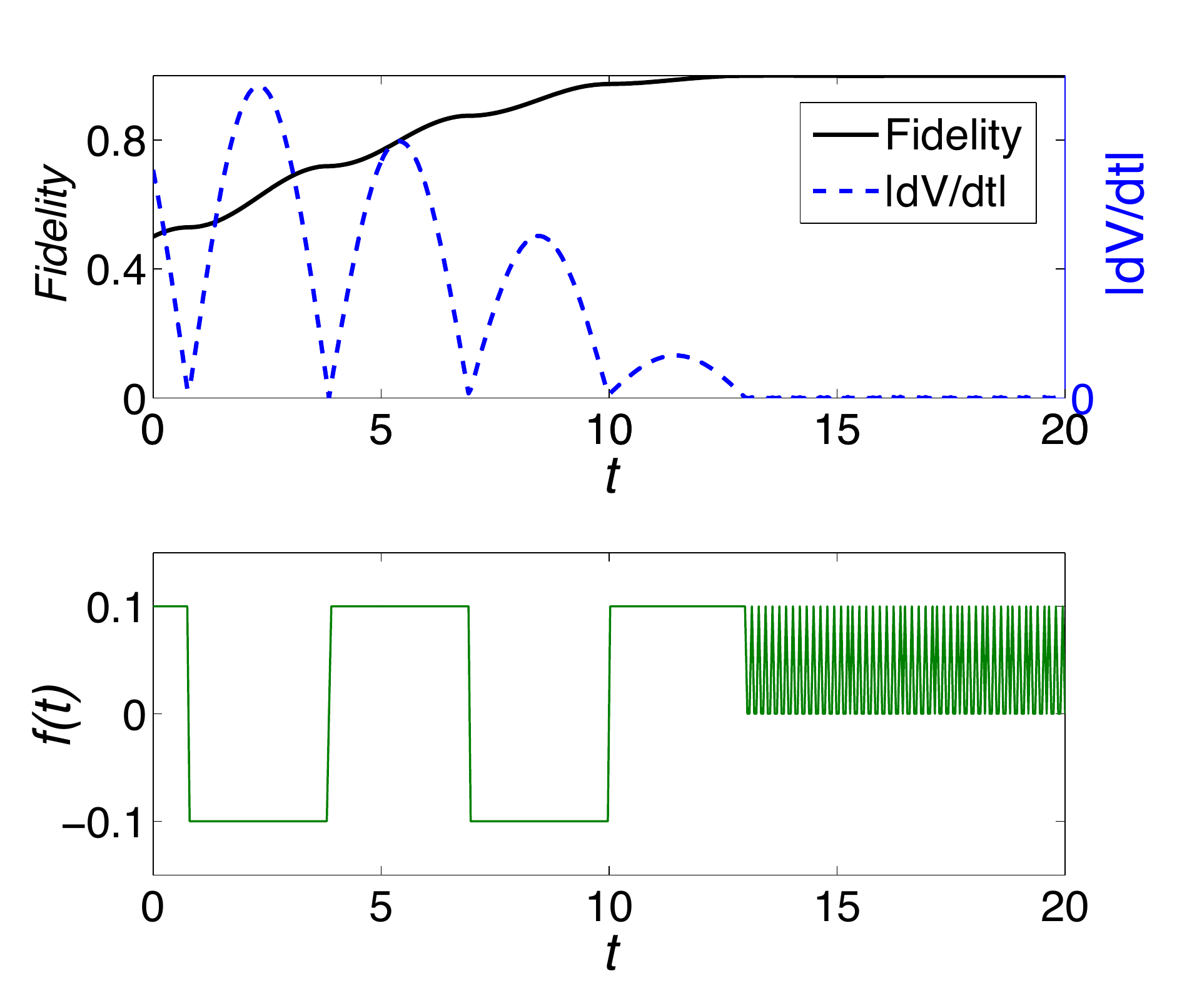}
\caption{Numerical simulations of the two-level system under the
Lyapunov control. $\omega=1$, $S=0.1$, the initial state is chosen
as $|\psi(0)\rangle=\frac 1{\sqrt
2}[|e\rangle+e^{-i\pi/4}|g\rangle]$.  } \label{fidelity}
\label{fig1simu}
\end{figure}

Before analytical calculations and analysis, we present a numerical
simulation for the two-level system under control.  The system
starts with an arbitrary state,
$|\psi_0\rangle=\cos\frac{\gamma_0}2|e\rangle+e^{i\phi}\sin\frac{\gamma_0}2|g\rangle$
with $\gamma_0\in(0,\pi)$ and $\phi\in[0,2\pi]$,  the target state
is $|e\rangle$. Fig. \ref{fig1simu} shows the evolution of the fidelity,  the control
field and the time-derivative of the Lyapunov function with time. In the simulation, we set $\gamma_0=\pi/2$,
$\phi=-\pi/4$, $\omega=1$ and $S=0.1$. Comparing the time derivative
$dV/dt$ and the control field $f(t)$ in Fig. \ref{fig1simu}, we
observe  that the control field changes at $dV/dt=0,$ or
$\text{Im}(ab^*)=0$ in Eq. (\ref{con_law}). Without loss of
generality, $a$ can be set to be a real number, and $\phi$ and  $\gamma_0$ determine
the design of the control field. In addition, we find that the
control field alters very quickly when the system is in the vicinity of
the target state.

\subsection{Dynamics evolution under control}

Assume that the initial state of a two-level system  is
\begin{eqnarray}
|\psi_0\rangle&=&\cos\frac{\gamma_0}2|e\rangle+
\sin\frac{\gamma_0}2e^{i\phi}|g\rangle \nonumber\\
                &\equiv&a_0|e\rangle+b_0|g\rangle , \label{inistate}
\end{eqnarray}
where $\gamma_0\in[0,\pi]$,  and $\phi\in[0, 2\pi]$ is
the relative phase. With different
parameters  $\gamma_0$ and $\phi$, $|\psi_{0}\rangle$ can represent an arbitrary
pure state (ignoring the global phase). Let the target state $|e\rangle$ correspond to the
north pole on the Bloch
sphere. Since $\text{Im}(a_0b_0^*) {=}{-}\frac {\sin\phi\sin\gamma_0} 2$, using the method in
(\ref{qubitcondition}), the first control field is calculated as,
\begin{eqnarray}
f=\left\{
\begin{array}{l}
S,\ {\text{Im}(a_0b_0^*)<}0\  (  0{<}\phi{<}\pi ), \\
0,\ {\text{Im}(a_0b_0^*)=}0\ (  \phi{=}0, \pi ), \\
{-}S,\ {\text{Im}(a_0b_0^*)>}0\ (  \pi{<}\phi{<}2\pi).
\end{array}
\right. \label{controllawphi}
\end{eqnarray}
Assume that this control would last until time $\tau$; i.e., the duration
of this control is $\tau$. With this control, the state evolves to
\begin{widetext}
\begin{eqnarray}\label{aftercontrol}
|\psi_{\tau}\rangle&=&[(e^{-iE_+\tau}\cos^2\frac{\theta}2
+e^{-iE_-\tau}\sin^2\frac{\theta}2 )\cos\frac{\gamma_0}2
+\frac 1 2(e^{-iE_+\tau}-e^{-iE_-\tau})
\sin\theta\sin\frac{\gamma_0}2e^{i\phi}]|e\rangle \nonumber\\
&+&[\frac 1 2(e^{-iE_+\tau}-e^{-iE_-\tau})\sin\theta \cos\frac{\gamma_0}2
+(e^{-iE_-\tau}\cos^2\frac{\theta}2+
e^{-iE_+\tau}\sin^2\frac{\theta}2)\sin\frac{\gamma_0}2 e^{i\phi}  ]|g\rangle \nonumber\\
&\equiv&a_{\tau}|e\rangle+b_{\tau}|g\rangle.
\end{eqnarray}
From
the design of the control law in (\ref{qubitcondition}),  we
find that a control field would last until $\text{Im}(a_{\tau}b^*_{\tau})$ changes
sign. Then $\tau$ can be given by solving
$\text{Im}(a_{\tau}b^*_{\tau})=0.$ Meanwhile, the sign of
$\text{Im}(a_{\tau}b^*_{\tau})$ determines the next  control field.
Simple algebra shows that
\begin{eqnarray}\label{switchcondition}
\text{Im}(a_{\tau}b^*_{\tau})&=&-\sin\phi\cdot
\frac 1 2\sin\gamma_0(\frac 1 2 +\frac 1 2\cos^2\theta)
\cdot\cos 2E_+\tau-\cos\phi
\cdot \frac 1 2\sin\gamma_0
\cdot\cos\theta\cdot\sin 2E_+\tau \nonumber\\
&-&\sin\phi\cdot \frac 1 2\sin\gamma_0
\cdot\sin^2\theta\cdot\frac 1 2 \cos 2E_+\tau
+\frac 1 2\cos\gamma_0\cdot\sin\theta\cdot\sin 2E_+\tau.
\end{eqnarray}
\end{widetext}
Since $a_{\tau}$ and $b_{\tau}$ are a function of  $\phi$, $\gamma_0$ and
$\theta$, the duration $\tau$ would be determined by these  three parameters.
Thereby, we arrange our discussions to cover the following cases.

(i) In the case of $\phi\neq 0$, the system would be steered from an
arbitrary state to a state on the $xz$ plane on the Bloch sphere,
i.e., $|\tilde{\psi}\rangle=\cos\frac{\gamma}2|e\rangle+
\sin\frac{\gamma}2|g\rangle.$  Note that the relative phase is zero,
i.e., $\tilde{\phi}=0$.

(ii) In the case of $\phi=0$ and $\gamma_0>\theta$, the control field switches slowly
between $f=S$ and $f=-S$ until $\gamma_0<\theta$.
Two situations, $\gamma_0\geq2\theta$ and  $\theta<\gamma_0<2\theta$, will be separately discussed.

(iii) In the case of $\phi=0$ and $0<\gamma_0\leq\theta$, the
control becomes inefficient. The control field  switches very quickly,
and the system may not be steered to the target state $|e\rangle$.

Given an initial state with $\phi\neq 0$ (i.e., case (i)), the first control would
steer the system into a state with zero relative phase, i.e., $\tilde{\phi}=0$.
The second control process  begins with the final state of the first control process (i.e., case (ii)).
Using the control law, a state with a zero relative phase yields zero control fields.
Hence, for a practical quantum system, a free evolution plays an important role at the beginning of the second control period.
It accumulates  a relative phase and triggers  the second control. Similar control processes would be
repeated until $0<\gamma_0\leq\theta$ (i.e., case (iii)).

For an arbitrary  initial state, the listed cases cover all situations
encountered in the method of optimal Lyapunov control  used in (\ref{qubitcondition}).
We will discuss these cases  in the next three subsections,
where the global phase of the quantum state is neglected throughout  the discussions.

\subsubsection{The case of $\phi\neq0$} \label{case1}

\begin{figure}
\includegraphics[width=0.8\columnwidth,height=0.5\columnwidth]{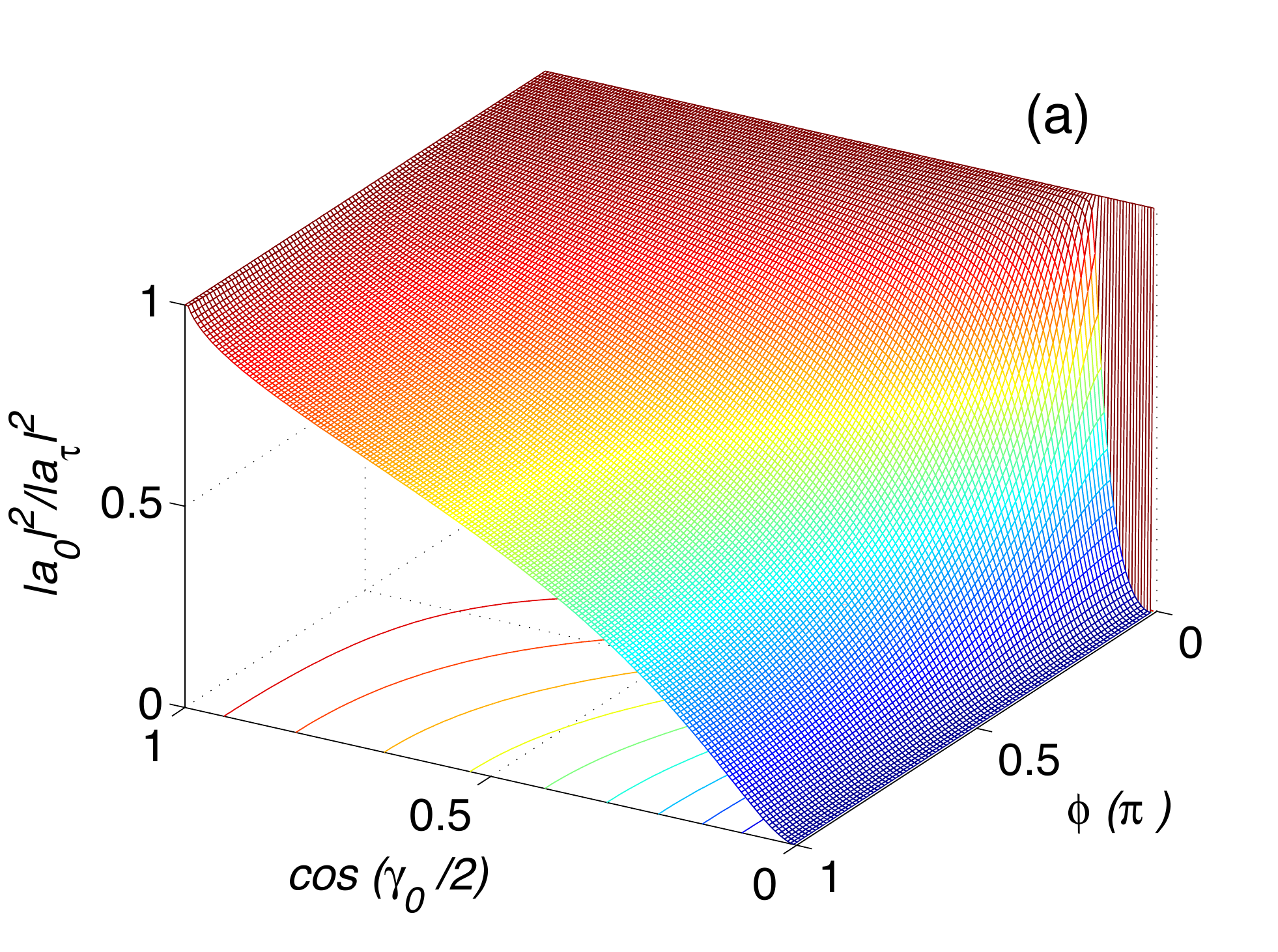}
\includegraphics[width=0.8\columnwidth,height=0.5\columnwidth]{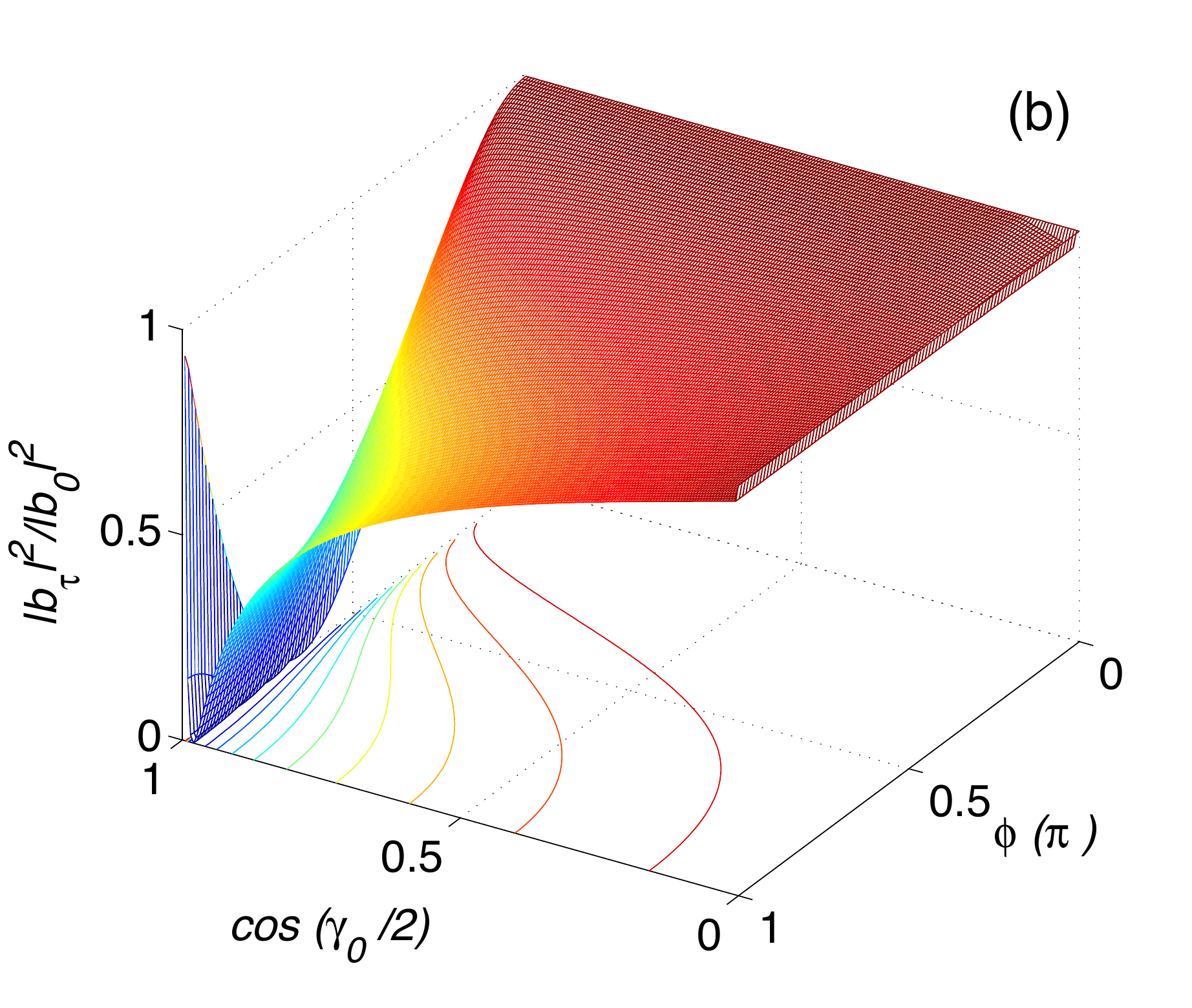}
\includegraphics[width=0.8\columnwidth,height=0.5\columnwidth]{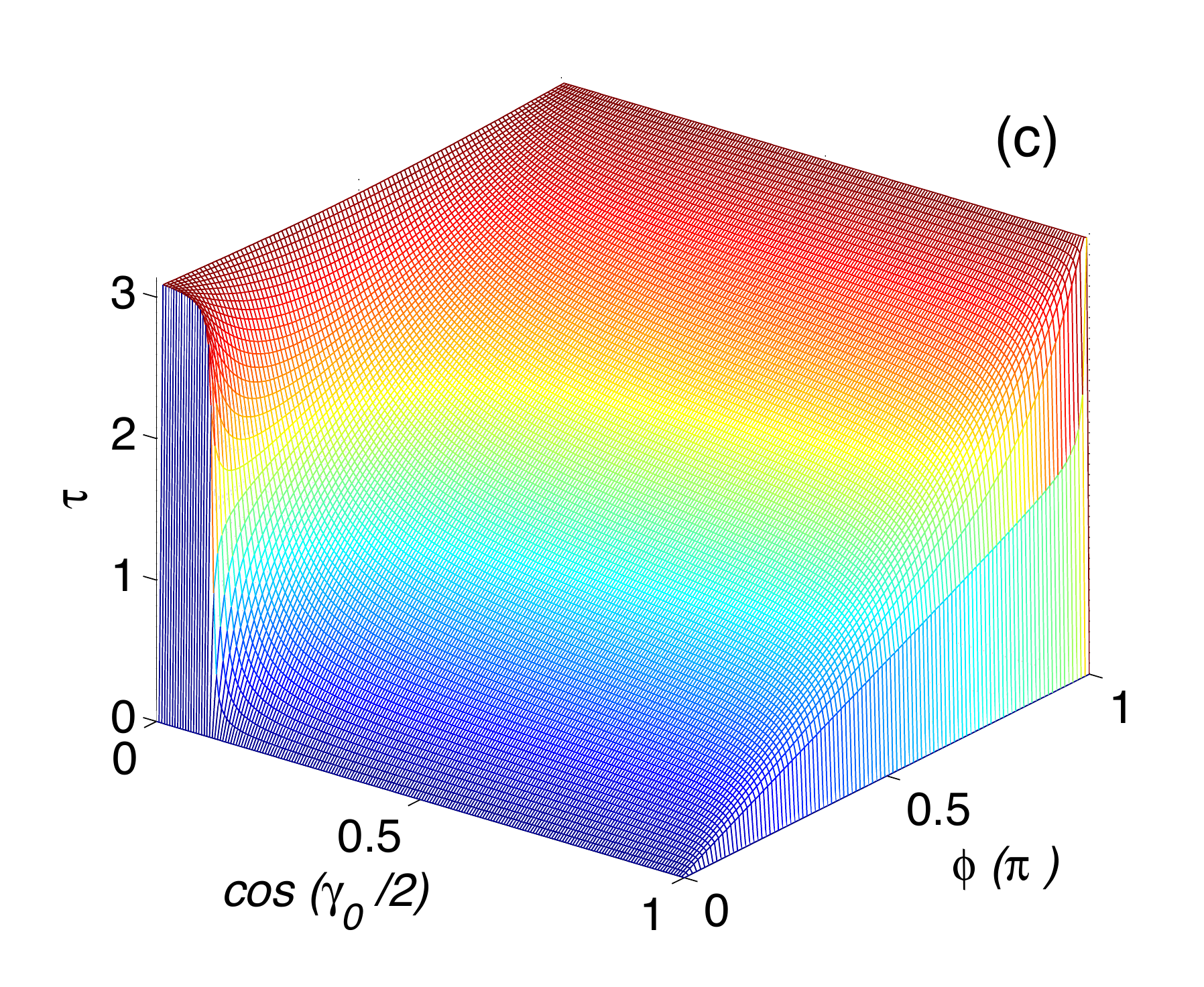}
\caption{The state after a control with duration $\tau$ versus the state
before the control, with $\omega=1$ and $S=0.1$.
(a) is plotted for  $|a_{0}|^2/|a_{\tau}|^2$, (b) for $|b_{\tau}|^2/|b_0|^2$,
(c) for the duration  $\tau$. }
\label{fig2}
\end{figure}

In this case, the control field is determined by
the sign of $\texttt{Im}(a_0b_0^*)$. The
duration  $\tau$ of this control process is determined by
$\texttt{Im}(a_{\tau}b^*_{\tau})=0$. From Eq. (\ref{switchcondition}), we have
\begin{eqnarray}
\tan 2E_+\tau=\frac{\sin\phi}{\sin\theta\cot\gamma_0-\cos\phi\cos\theta}.
\end{eqnarray}
After this control process, the relative phase $\phi$
vanishes. In other words, with an arbitrary state as the
initial state, the control would fall into either case (ii) or (iii)
after the first control period. Population changes of the  system on the
two levels after the control can be described by
 $|a_0|^2/|a_{\tau}|^2$ or $|b_{\tau}|^2/|b_0|^2$,
where $a_{\tau}$ and $b_{\tau}$ are given in
Eq. (\ref{aftercontrol}). $|a_0|^2/|a_{\tau}|^2$ and
$|b_{\tau}|^2/|b_0|^2$ versus $\phi$ and $\gamma_0$ are shown in
Fig. \ref{fig2}-(a) and Fig.\ref{fig2}-(b), and the control duration
$\tau$ is shown in Fig. \ref{fig2}-(c). Fig. \ref{fig2} shows that, after
the first control process, the amplitude of $|e\rangle$   increases,
while the amplitude of $|g\rangle$  decreases.

If the initial state can be written in the following form,
\begin{eqnarray}
|\psi_i\rangle&=&\frac 12 [e^{iE_-\tau_i}
+e^{iE_+\tau_i}+(e^{iE_+\tau_i}-e^{iE_-\tau_i})\cos\theta]|e\rangle \nonumber\\
&+&\frac 1 2(e^{iE_+\tau_i}-e^{iE_-\tau_i})\sin\theta|g\rangle,
\end{eqnarray}
with $\forall\tau_i\in[0, {\pi}/{2E_+}]$, the state after the control would be
\begin{eqnarray}
U(\tau_i)|\psi_i\rangle=|e\rangle\langle e|,
\end{eqnarray}
i.e., the state can be  controlled to the target state by a single
control. The states that can be steered to the target state by a
single control are shown in Fig. \ref{fig2}-(b) (i.e., the points $(\cos(\gamma_0/2),\phi)$ that
satisfy $|b_{\tau}|^2/|b_0|^2=0$). A clear
demonstration  will be presented in Fig. \ref{fig7}-(a) and Fig.
\ref{fig7}-(b)).  This result will  be used to
propose an extended technique for optimal Lyapunov quantum
control in Sec. IV.

\subsubsection{The case of $\phi=0$ and $\gamma_0 > \theta$}\label{case2}
For a state with  $\phi=0$, we have
\begin{eqnarray}
|\psi_1\rangle_{\pm}=\cos\frac{\gamma_0}2|e\rangle \pm
\sin\frac{\gamma_0}2|g\rangle, \ \ \gamma_0 > \theta.
\end{eqnarray}
This state may be the resulting state of the first control process.

\begin{figure}
\includegraphics[width=0.5\columnwidth,height=0.54\columnwidth]{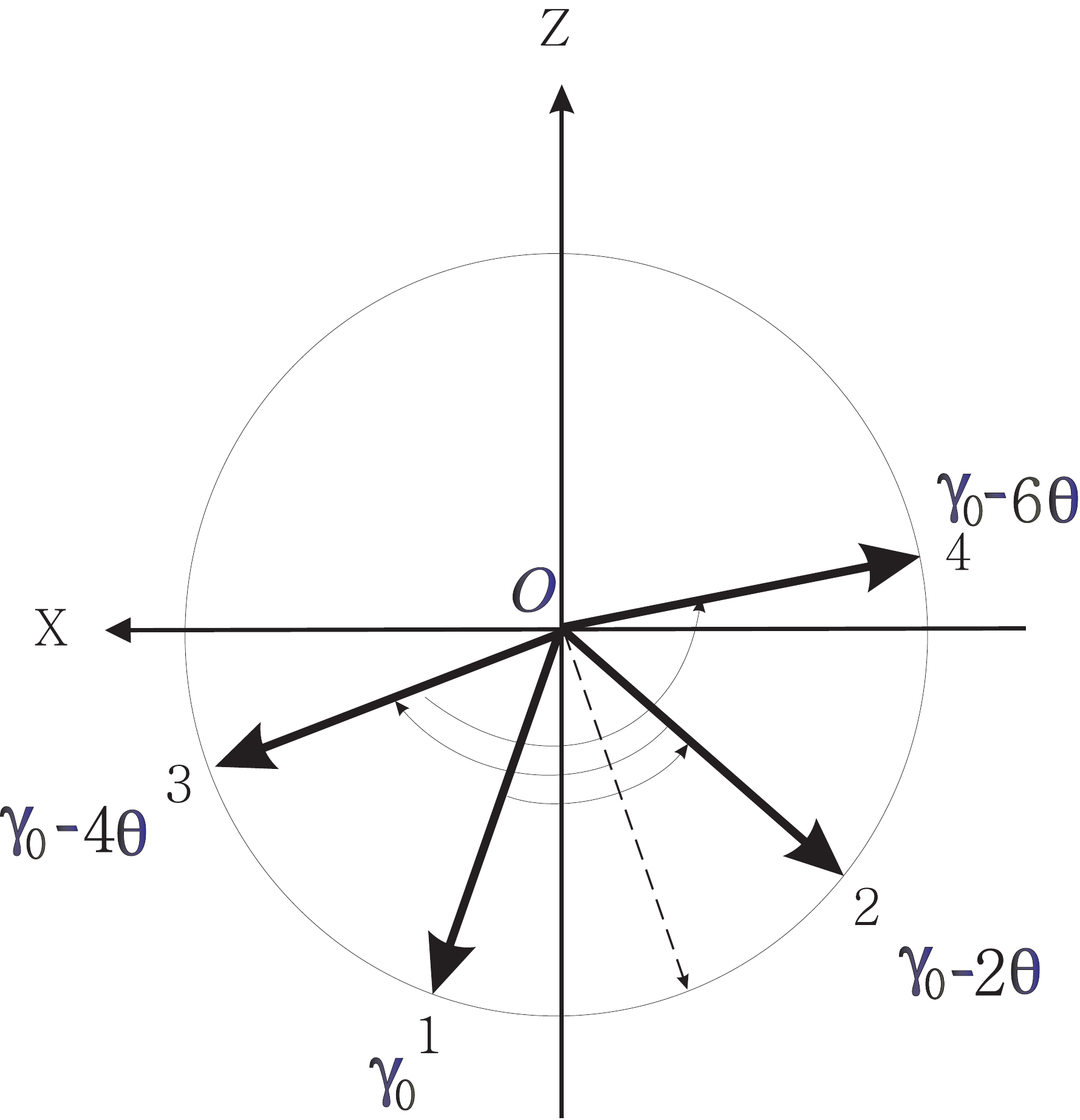}
\caption{Illustration of states after controls on the Bloch sphere. This
illustration is for the  case of $\gamma_0\geq2\theta$. Each control
brings the Bloch vector one step closer to the target by reducing
the azimuthal angle. $1(\gamma_0)\rightarrow2(\gamma_0-2\theta)
\rightarrow3(\gamma_0-4\theta)\rightarrow4(\gamma_0-6\theta)\rightarrow\cdots$.}
\label{fig3}
\end{figure}

\begin{figure}
\includegraphics[width=0.5\columnwidth,height=0.56\columnwidth]{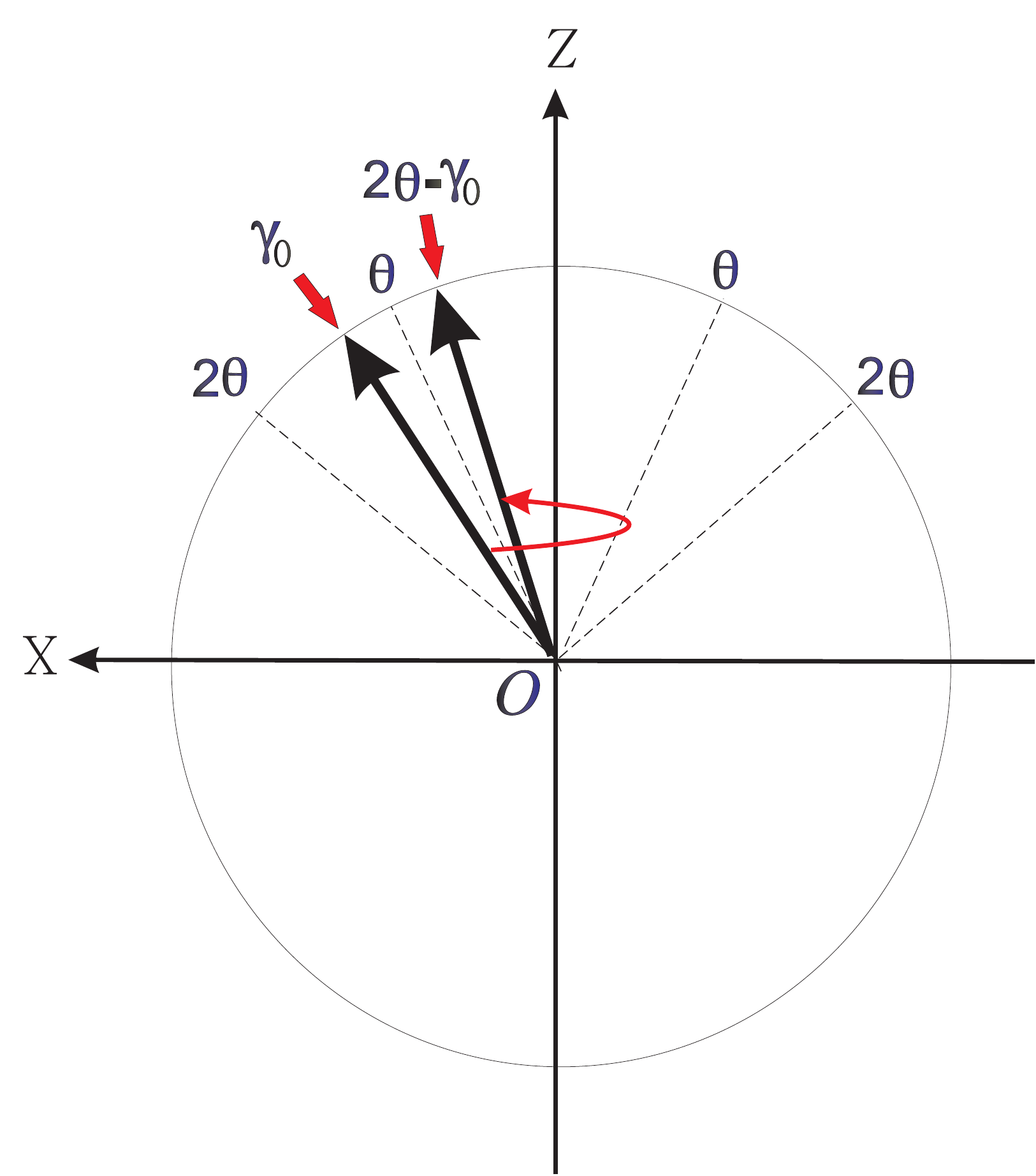}
\includegraphics[width=0.5\columnwidth,height=0.56\columnwidth]{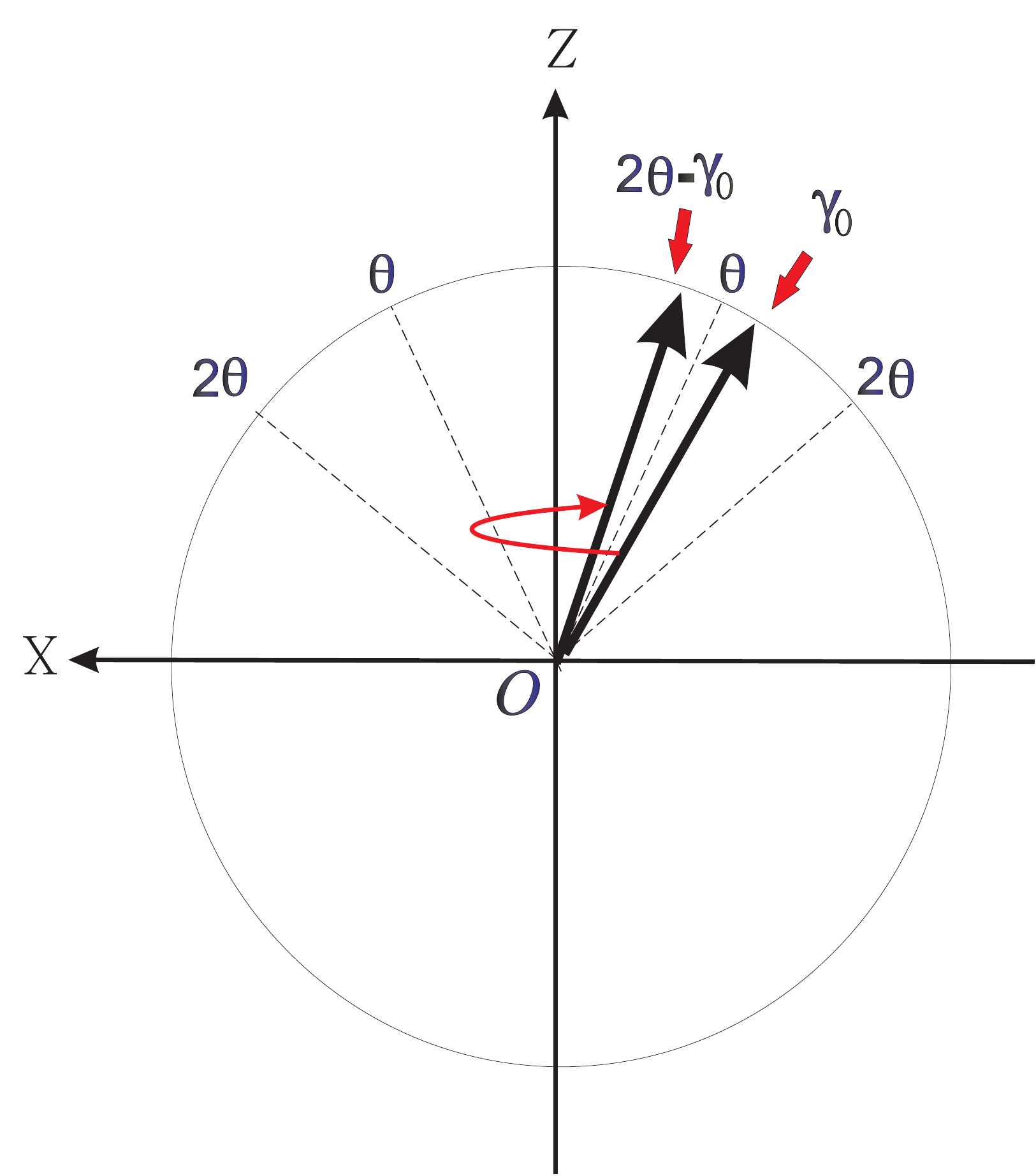}
\caption{ Illustration of Bloch vectors before and after the control
for the case  $\phi=0$ and $\theta<\gamma_0<2\theta$. (a) is for
initial state $|\psi_1''\rangle_+$,   while (b) for initial state
$|\psi_1''\rangle_-$. } \label{fig4}
\end{figure}

Note   that  at this moment $\text{Im}(a_{\tau}b^*_{\tau})=0$,
and the control field  satisfies  $f=0$. However,
a practical system will acquire a relative phase $\phi$ in an extremely short time due to the free evolution, which would trigger the
control. To be specific, assume that the free evolution time is  $\delta
t'\rightarrow 0$. The state after this free evolution is
\begin{eqnarray}
|\psi_1'\rangle_{\pm}=e^{-\frac{i\omega \delta t'}2}\cos\frac{\gamma_0}2 |e\rangle
                    \pm e^{\frac{i\omega \delta t'}2}\sin\frac{\gamma_0}2 |g\rangle.
\end{eqnarray}
It turns out that $\text{Im}(a'_{\pm}b'^{*}_{\pm})=
\mp \cos\frac{\gamma_0}2\sin\frac{\gamma_0}2\sin{\omega \delta t}$.
This triggers  a control process with a control field $f=\pm S $.
With this control, after $\delta t''$, we
have
\begin{eqnarray} \label{case2condition}
\texttt{Im}(a_{\pm}b_{\pm}^*) &\simeq& \mp \sin\omega
\delta t'\cdot \frac 1 2\sin\gamma_0\cdot\cos 2E^+\delta t'' \nonumber\\
&\mp & \sin(\gamma_0-\theta)\sin 2E^+\delta t''.
\end{eqnarray}
Here, $\cos\omega \delta t'\simeq 1$ has been used. We find that
$\texttt{Im}(a_{+}b_{+}^*)<0$ (and $\texttt{Im}(a_{-}b_{-}^*)>0$)
is kept, since $\gamma_0>\theta$. In fact, the  control field $f=S$ or
$f=-S$ will last  for  a period of $\tau_1$ determined by
\begin{eqnarray}
\mp \sin 2E_+\tau_1 \cdot \frac 1 2\sin(\theta-\gamma_0)=0.
\end{eqnarray}
Hence, $\tau_1=\pi/2E_+$. This control process steers the system
from $|\psi_1\rangle_{\pm}$ to a state
\begin{eqnarray} \label{case2control}
|\psi_2\rangle_{\pm}&=&[(e^{-iE_+\tau_1}
\cos^2\frac{\theta}2+e^{-iE_-\tau_1}\sin^2\frac{\theta}2 )\cos\frac{\gamma_0}2 \nonumber\\
&+&\frac 1 2(e^{-iE_+\tau_1}-e^{-iE_-\tau_1})
\sin\theta\sin\frac{\gamma_0}2]|e\rangle \nonumber\\
&\pm &[\frac 1 2(e^{-iE_+\tau_1}-e^{-iE_-\tau_1})
\sin\theta \cos\frac{\gamma_0}2 \nonumber\\
&+&(e^{-iE_-\tau_1}\cos^2\frac{\theta}2+e^{-iE_+\tau_1}
\sin^2\frac{\theta}2)\sin\frac{\gamma_0}2  ]|g\rangle \nonumber\\
&=&-i[\cos(\frac{\gamma_0}2-\theta)|e\rangle \mp \sin(\frac{\gamma_0}2-\theta)|g\rangle].
\end{eqnarray}
Eq. (\ref{case2control}) shows that such a control brings the state closer to
the target state. In the Bloch sphere representation,
this control reduces the azimuthal angle
by   $2\theta$, taking  the Bloch vector one step closer to the positive $z$ axis
(see Fig. \ref{fig3}). The same control would be repeated. After
$n$ times,  the state  evolves to
\begin{eqnarray}\label{l2theata}
|\psi_n\rangle_{\pm}&=&\cos(\frac{\gamma_0}2-n\theta)
|e\rangle+(-1)^n\sin(\frac{\gamma_0}2-n\theta)|g\rangle, \nonumber\\
&\equiv& \cos\frac{\gamma'}2|e\rangle+(-1)^n\sin\frac{\gamma'}2|g\rangle.
\end{eqnarray}
This type of control continues UNTIL
$\frac{\gamma_0}2-n\theta=\frac{\gamma}2 '<\theta$. Next, if
$0<\gamma'\leq\theta$, this type of control will stop. If
$\theta<\gamma'<2\theta$, another control would steer the system to the regime
$0<\gamma'\leq\theta$.
Recalling Eq. (\ref{l2theata}), after $n$ control processes, if  a state with
$\gamma_0>2\theta$ falls into the regime
$\theta < \gamma_0 < 2\theta$,
\begin{eqnarray}
|\psi_1''\rangle_{\pm}=\cos\frac{\gamma_0}2|e\rangle \pm
\sin\frac{\gamma_0}2|g\rangle, \theta < \gamma_0 < 2\theta,
\end{eqnarray}
we can employ the same analysis as that in the case of $\gamma_0>2\theta$ to find a control that will steer the initial state $|\psi_1''\rangle_{\pm}$ to
\begin{eqnarray}
|\psi_2''\rangle_{\pm} &=&
e^{i\epsilon}\Big[\cos\frac{|2\theta-\gamma_0|}2|e\rangle
\pm \sin\frac{|2\theta-\gamma_0|}2|g\rangle \Big]  \nonumber \\
&=&\cos\frac{\gamma}2''|e\rangle \pm \sin\frac{\gamma}2''|g\rangle.
\label{fa_theata}
\end{eqnarray}
It is clear that $0<|2\theta-\gamma_0|<\theta$ since
$\theta<\gamma_0<2\theta$. Namely, the states finally fall into the
regime $0<\gamma_0\leq\theta$ (see Fig. \ref{fig4}).
Hence, using the second type of control processes, the state can be driven to
\begin{eqnarray}\label{finalstate-of-second-process}
|\psi'_n\rangle_{\pm}=\cos\frac{\gamma}2'|e\rangle \pm
\sin\frac{\gamma}2'|g\rangle,\ ( 0<\gamma'\leq \theta ).
\end{eqnarray}

If
\begin{eqnarray}\label{exactly-controllable}
\frac{\gamma_0}2-n\theta=0, (n=1,2,...),
\end{eqnarray}
the system can be steered exactly to the target state by $n$
controls. The number of controls $n$  can be given  by
\begin{eqnarray}
\theta=\frac{\gamma_0}{2n} \ \ \
\Leftrightarrow  \ \ \
\arctan\frac{2f}{\omega}=\frac{\gamma_0}{2n}.
\label{case2inf}
\end{eqnarray}
On the other hand, given a number of  controls $n_0$, we can
calculate the control strength such that the
system can be steered exactly to the target state by $n_0$ times of control processes

When (\ref{exactly-controllable}) is not satisfied, we consider a new control process for the state in (\ref{finalstate-of-second-process})
in the next subsection.



\subsubsection{The case of $\phi=0$ and $0<\gamma_0 \leq \theta$ } \label{case4}

Now we discuss the  control starting with a state,
\begin{eqnarray}
|\psi_1\rangle_{\pm}=\cos\frac{\gamma_0}2|e\rangle
\pm \sin\frac{\gamma_0}2|g\rangle, \ \ 0<\gamma_0\leq\theta.
\end{eqnarray}
The control field is zero at the beginning by the control law. After an infinitesimal
$\delta t'\rightarrow0$ free evolution,
a control with control field  $f=\pm S$ is triggered.  After a period of  $\delta t''$,
we have,
\begin{eqnarray}
\texttt{Im}(a_{\pm}b_{\pm}^*)
&=&\pm \frac 1 2\sin\omega \delta t'\cdot
\sin\gamma_0\cdot\cos 2E^+\delta t''\nonumber\\
&\mp & \sin(\gamma_0-\theta)\sin 2E^+\delta t''.
\end{eqnarray}

A careful
examination shows that this   is  different from Eq.
(\ref{case2condition}), because the sign of
$\texttt{Im}(a_{\pm}b_{\pm}^*)$ in Eq. (\ref{case2condition}) does
not change after the infinitesimal control, and  the control would
last until the azimuthal angle lost $2\theta$, whereas in the case
of $\gamma_0<\theta$, the control can not last for such a long time
since $\sin(\gamma_0-\theta)<0$. In fact, the control field
switches very quickly in this case.  The (infinitesimal) duration
of the control $\delta t''$ satisfies,
\begin{eqnarray}
\tan 2E_+\delta t''=\frac{\sin\omega \delta t'}
{\sin\theta\cot\gamma_0-\cos\omega \delta t'\cos\theta}.
\end{eqnarray}
After this duration, the system evolves to
$|\psi\rangle=\cos\frac{\gamma}2'|e\rangle
+\frac{\gamma}2'|g\rangle$ with $0<\gamma'\leq\theta$ again.
Since the control time $\delta t''$ are determined by the free-evolution
time $\delta t'$, therefore,  $\delta t'$  can not be ignored for a practical system.

For a very small duration, the control
could drive the system closer to the target, since
\begin{eqnarray}\label{eq39}
\frac{|\cos\frac{\gamma}2'|^2}{|\cos\frac{\gamma_0}2|^2}
&\simeq & 1+\frac{(\omega \delta t')^2\sin^2
\frac{\gamma_0}2\sin\theta [\sin(\theta-\gamma_0)
+\frac {\sin\gamma_0}2]}{\sin^2(\theta-\gamma_0)} \nonumber\\
&=&1+A(\gamma_0)\delta t'^2 > 1,
\end{eqnarray}
where we have neglected the high order terms of $\delta t'$.
Eq. (\ref{eq39}) tells us that in this infinitesimal control,
the convergence of the system towards the target state could depend on the
free evolution time $\delta t'$. Hence, an additional free evolution
may help improve the effectiveness of the control law in (\ref{qubitcondition}).

\subsection{Control limit and  strength of control fields} \label{controldisscuss}
As mentioned above,  the state of the system can be described by $\gamma_0$ and
$\phi$. When $\gamma_0$ satisfies $0<\gamma_0\leq\theta$, the control becomes
inefficient, i.e., the control fields switch very quickly but the system may not evolve  towards the target. We refer to this control as
{\it fast-switching control (FSC)}, and the controls before this as {\it slow switching
controls (SSC)}. With this knowledge,  one may wonder, if we can stop the control before
$\gamma_0$ enters the   regime $0<\gamma_0\leq\theta$. To answer this question,
we examine the fidelity achieved by the slow-switching controls.

\begin{figure}
  \includegraphics[width=0.75\columnwidth,height=0.5\columnwidth]{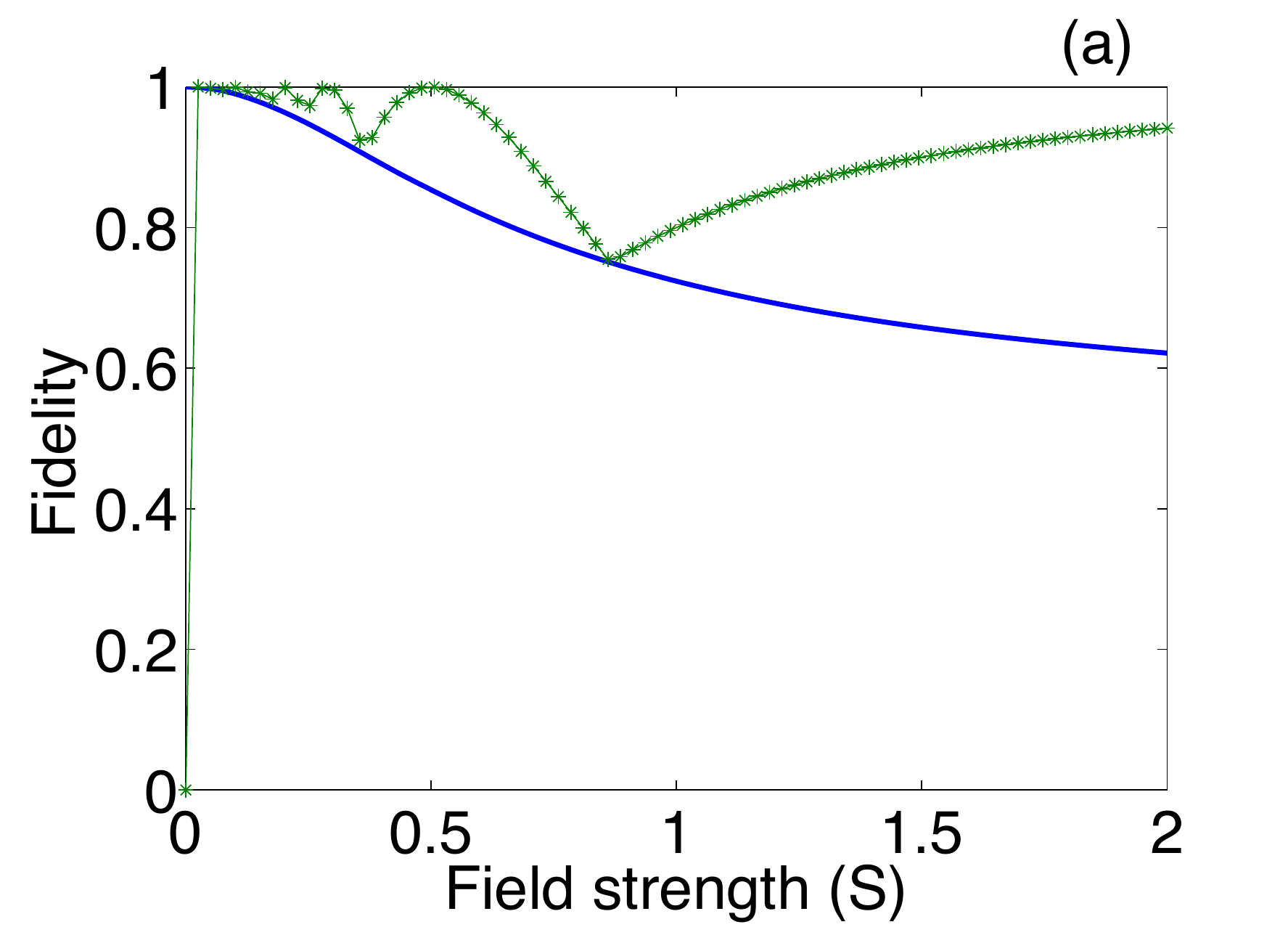}
  \includegraphics[width=0.8\columnwidth,height=0.5\columnwidth]
  {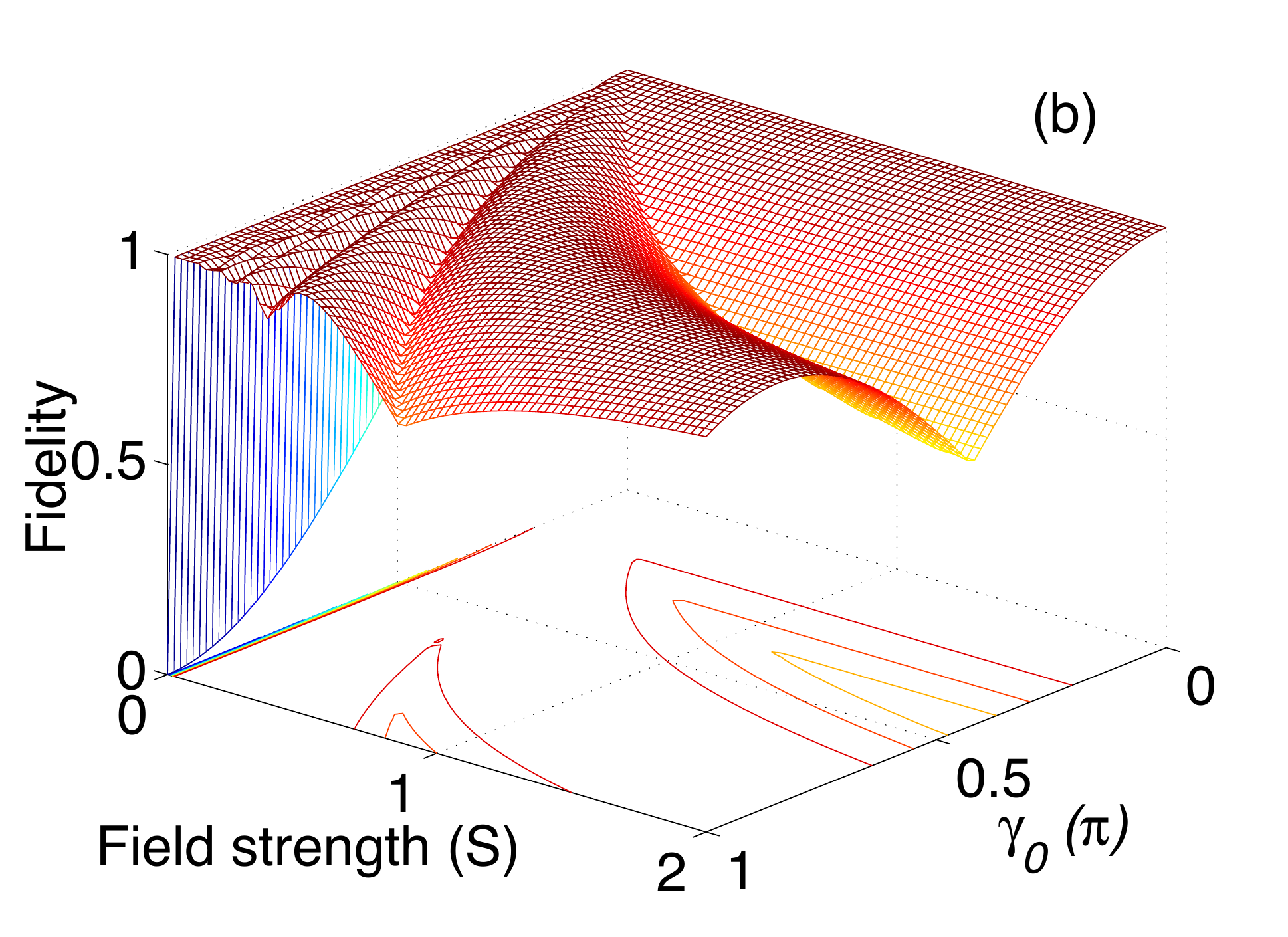}
  \caption{ The fidelity of SSC versus the strength of the control field.
The solid line in (a) plots the the lower bound of  the fidelity reached by SSC,
given in Eq. (\ref{limF}), while the dot-solid line is for the fidelity of SSC
with initial state, $1/\sqrt{2}(|e\rangle+|g\rangle)$.
Figure (b) shows the fidelity of SSC versus  the strength
of the control fields and initial states, $|\psi(0)\rangle=\cos\frac{\gamma}2|e\rangle
+\sin\frac{\gamma}2|g\rangle$, $\gamma \in[0,\pi]$.}
\label{fig5}
\end{figure}

Denoting  $\gamma_f$  the  azimuthal angle reached by the SSC,
we find that
\begin{eqnarray}
F=\cos^2\frac{\gamma_f}2\geq \cos^2\frac{\theta}2=
\frac 1 2+\frac 1{2\sqrt{1+(2S/\omega)^2}}. \label{limF}
\end{eqnarray}
Fig. \ref{fig5} shows the fidelity of the SSC versus the strength of the control field and
the initial state.
We find from Fig. \ref{fig5} that the smaller the strength of the
control field is, the larger the fidelity of the SSC could be. Note that
small control strength needs more alternations  (e.g., from $-S$ to
$S$) in the control fields.

\begin{figure}
\includegraphics[width=0.8\columnwidth,height=0.5\columnwidth]{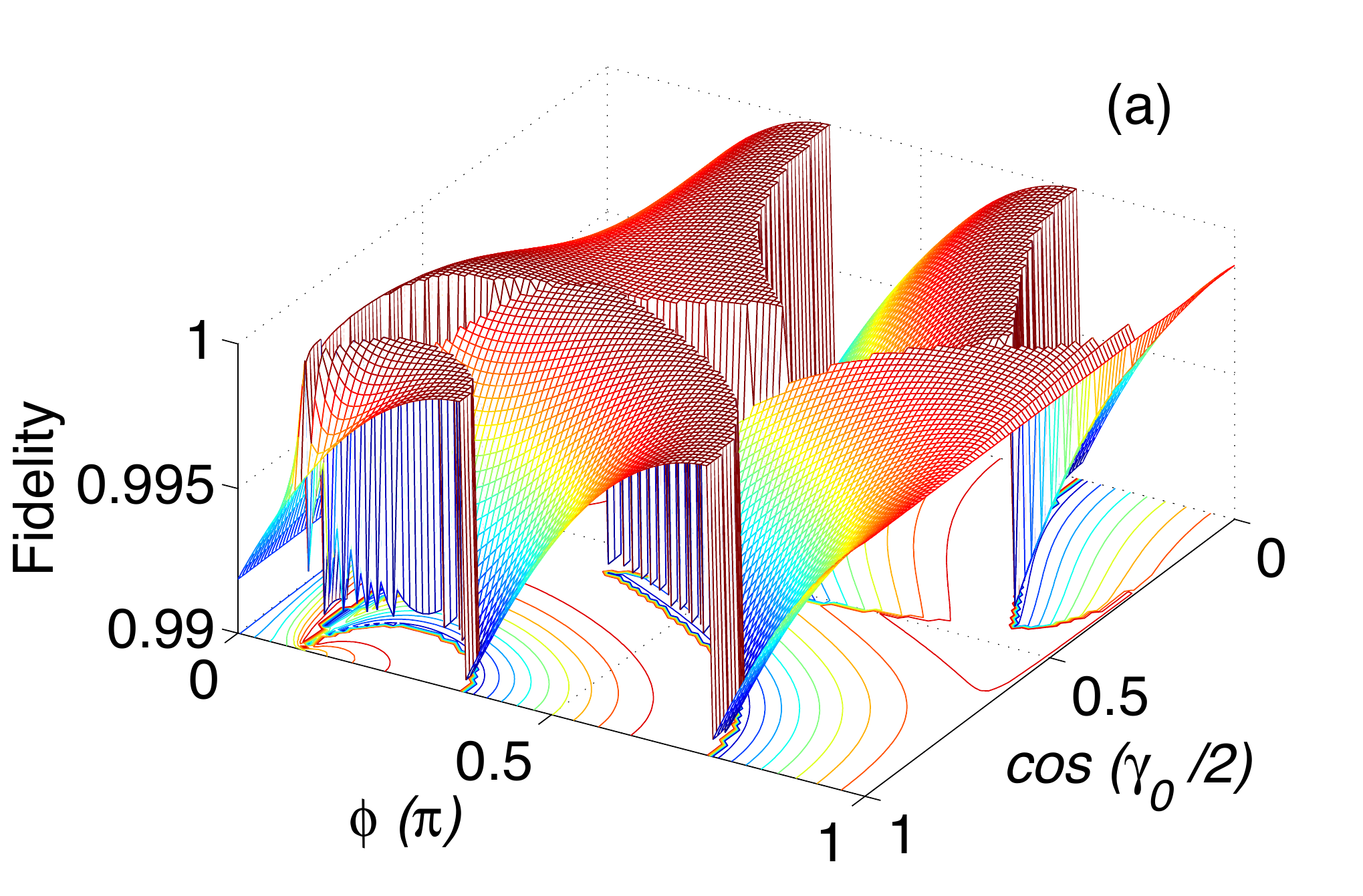}
\includegraphics[width=0.82\columnwidth,height=0.5\columnwidth]{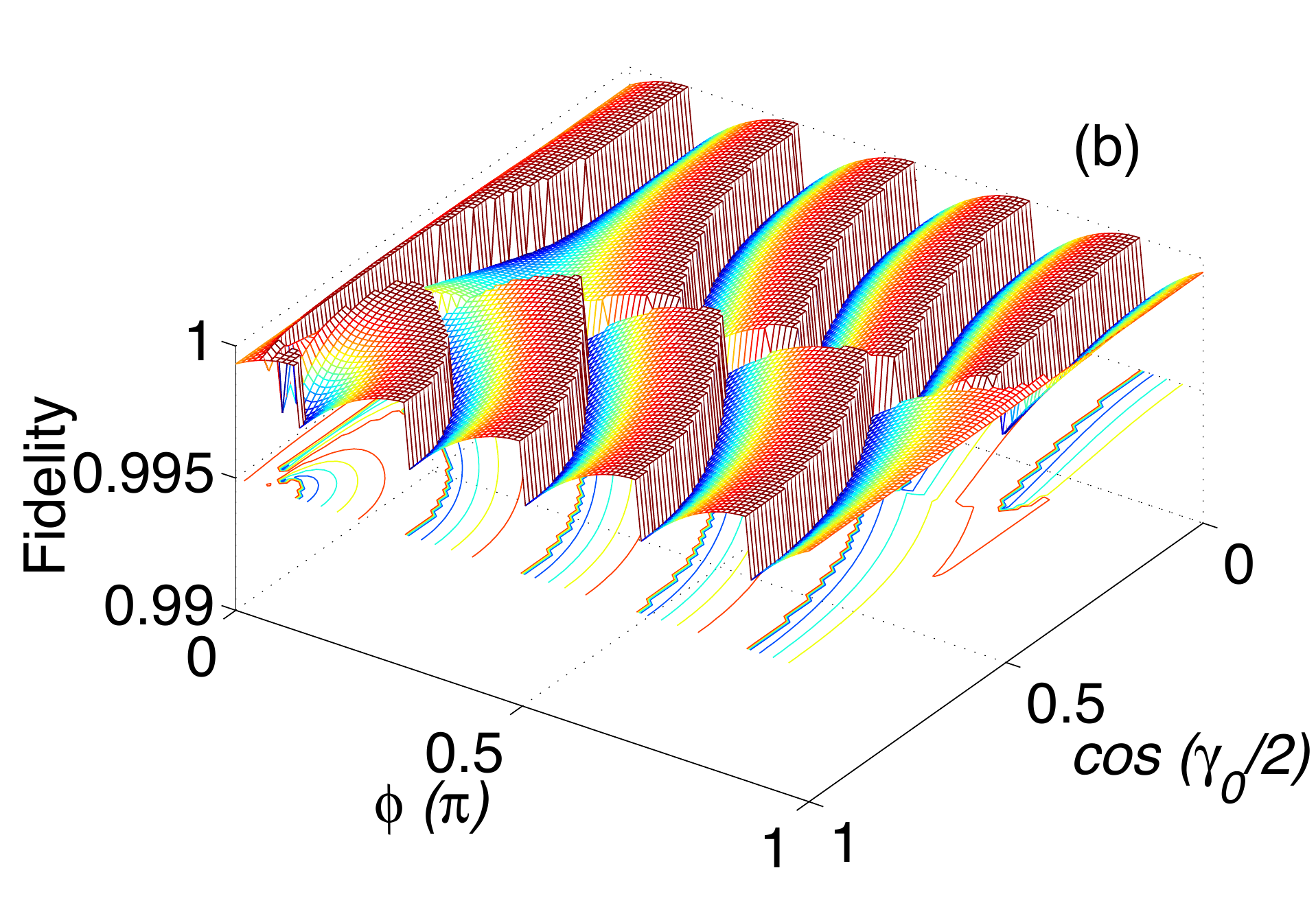}
\caption{Fidelity of SSC as a function of initial states,
$|\psi(0)\rangle=\cos\frac{\gamma_0}2|e\rangle+e^{i\phi}\sin\frac{\gamma_0}2|g\rangle$,
(a) and (b) are for different control strengths,  (a) $S=0.1$; (b) $S=0.05$.}
\label{fig6}
\end{figure}

For an arbitrary initial state,
$|\psi(0)\rangle=\cos\frac{\gamma}2|e\rangle+e^{i\phi}\sin\frac{\gamma}2|g\rangle$,
the resulting  state of the system after SSC becomes,
\begin{eqnarray}
|\psi_f\rangle=
\left\{
\begin{array}{l}
\cos\frac{2\theta-\gamma'}2|e\rangle+\sin\frac{2\theta-\gamma'}
2|g\rangle, \  ( \theta  < \gamma' \leq 2\theta), \\
\cos\frac{\gamma'}2|e\rangle+\sin\frac{\gamma'}2|g\rangle, \  (0<\gamma' \leq \theta), \\
\end{array}
\right.
\end{eqnarray}
where,
\begin{eqnarray}
&&\gamma'=\gamma-2n_{max}\theta,  \nonumber\\
&&\cos\frac{\gamma}2 = (e^{-iE_+\tau}\cos^2\frac{\theta}2
  + e^{-iE_-\tau}\sin^2\frac{\theta}2 )\cos\frac{\gamma_0}2  \nonumber\\
&& \ \ \ \ \ \ \ \   +\frac 1 2(e^{-iE_+\tau}-e^{-iE_-\tau})
\sin\theta\sin\frac{\gamma_0}2 e^{i\phi}, \nonumber \\
&& \tan 2E_+\tau = \frac{\sin\phi}{\sin\theta\cot\gamma_0-\cos\phi\cos\theta}.
\end{eqnarray}
Here, $n_{max}$ denotes the number of SSC which depends not only on the
initial states but also on the strength of the control. With these resulting
states, we can calculate the fidelity reached by the SSC.
Stronger control fields usually lead to
smaller fidelity of SSC and less number of controls (see Fig. \ref{fig6}).
From Fig. \ref{fig6}, we also find that with fixed strength of
the control field, the SSC itself
can drive some   initial states into the target state precisely.
Alternatively, for an specific initial state,
we can find a  control strength that
steers it to the target with fidelity one.


\section{Extended technique}

As discussed in the last section, the control law in (\ref{qubitcondition}) may become inefficient when
the system is very close to the target state. If we stop
the controls before the inefficient FSC, the
control fidelity can not reach a desirable value.
 In this section, we propose an extended
technique to improve the optimal Lyapunov quantum control in \cite{hou2012}.

Recall that an additional free evolution
may enhance the efficiency of the control. We combine
free evolution and external control into the extended technique where
different controls are used for a state with $\texttt{Im}(ab^*)=0$
according to whether $|a|^2<\cos^2\theta$ or not.
For a state close to the target ($|a|^2>\cos^2\theta$), we implement an additional
free evolution with $\Delta t$, while for a state far
from the target, we take  the same control law in (\ref{qubitcondition}).
Suitable design of $\Delta t$  can steer the system to the target state.

In Subsection III.B,
we  find that if an initial state satisfies,
\begin{eqnarray}
|\psi_i\rangle&=&\frac 12 [e^{iE_-t}+e^{iE_+t}
+(e^{iE_+t}-e^{iE_-t})\cos\theta]|e\rangle \nonumber\\
&+&\frac 1 2(e^{iE_+t}-e^{iE_-t})\sin\theta|g\rangle,
\label{ini_sc}
\end{eqnarray}
with parameter  $t$ in  $[0, {\pi}/{2E_+}]$, the state can be
steered into the target state by a single control.

To see clearly
what type of states can be steered to the target by a single control,
we rewrite
the state in Eq. (\ref{ini_sc}) in the following form,
\begin{eqnarray}
|\psi_i\rangle{=}e^{i\pi}\Big[e^{i\phi'}\sqrt{1{-}
\sin^2E_+t\sin^2\theta}|e\rangle+\sin E_+t\sin\theta|g\rangle\Big], \nonumber\\
\end{eqnarray}
where $\phi'$ is defined by
\begin{equation}
\tan\phi'=\frac{\cos E_+t}{\sin E_+t\cos\theta}. \label{phi}
\end{equation}
\begin{figure}
\includegraphics[width=0.75\columnwidth,height=0.5\columnwidth]{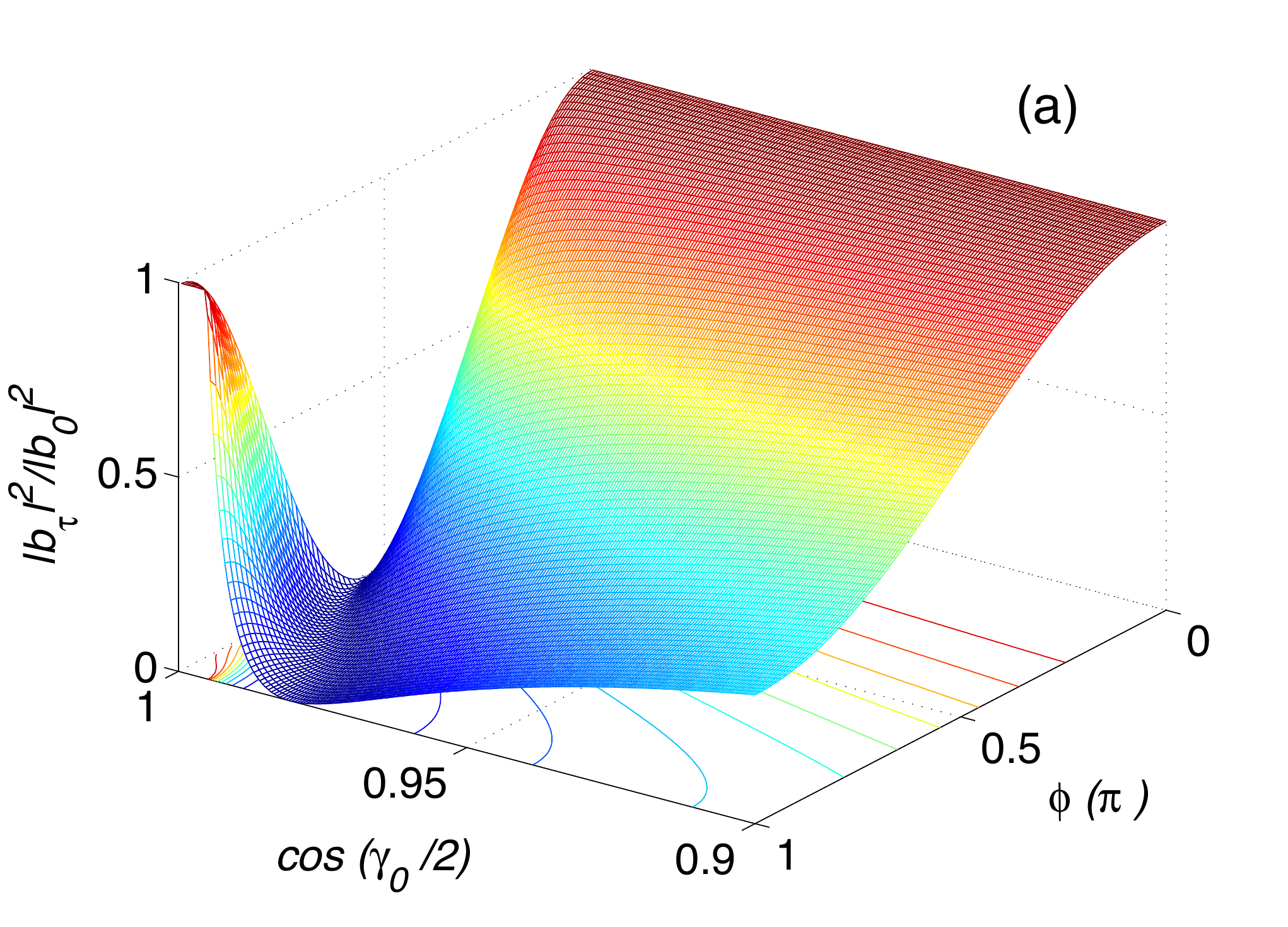}
\includegraphics[width=0.75\columnwidth,height=0.4\columnwidth]{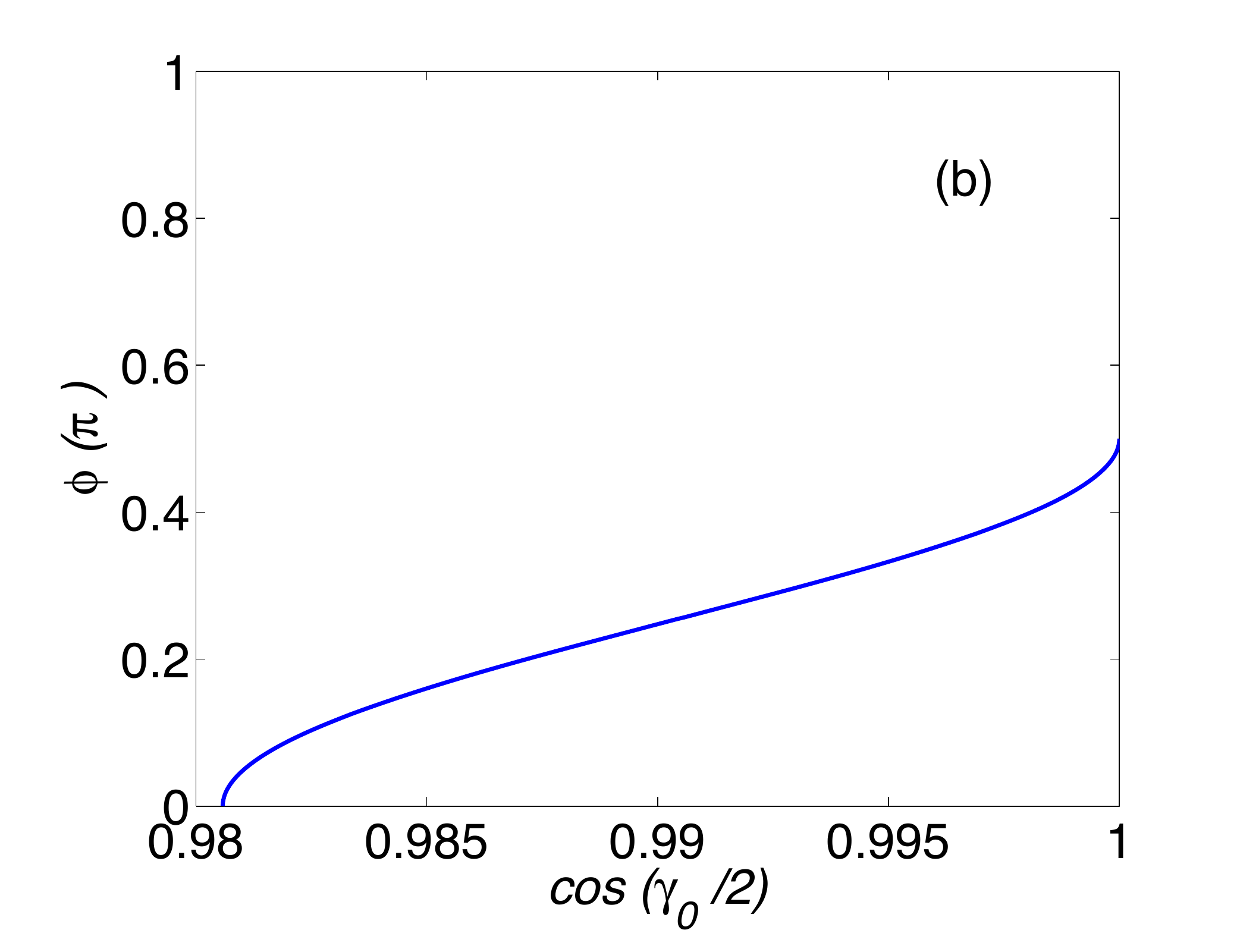}
\includegraphics[width=0.75\columnwidth,height=0.4\columnwidth]{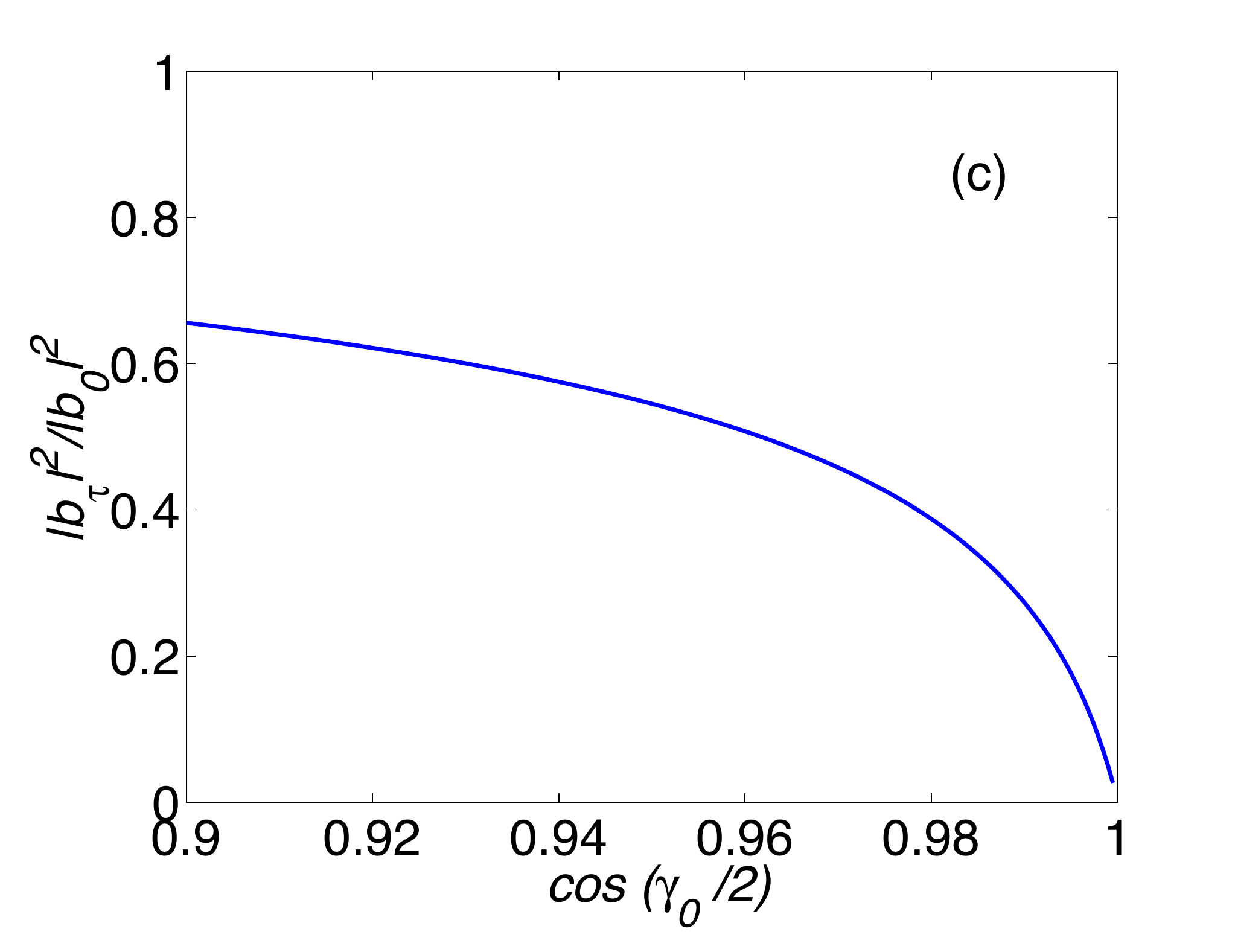}
\caption{ Figure (a) is an enlarged part of Fig.\ref{fig2}-(b).
(b) shows the states ( characterized by $\gamma$ and $\phi$ ) that
can be steered to the target by a single control.
(c) $|b_{\tau}|^2/|b_0|^2$ versus $\cos\frac{\gamma}2$ with  $\phi=\pi/2$ in
figure (a).}
\label{fig7}
\end{figure}
Since $\cos^2\theta\leq 1-\sin^2E_+t\sin^2\theta \leq 1$, and $1-\sin^2E_+t\sin^2\theta$
represents the probability to obtain the target state $|e\rangle$ when making a measurement,
any state $a|e\rangle+b|g\rangle$ with $|a|^{2}>\cos^2\theta$
can be controlled to the target state by a single control, provided that $\phi'$ satisfies the condition
in (\ref{phi}). It is worth mentioning that $\phi'$ is the relative phase, which can be
manipulated by changing the   free evolution time to any required value.

As shown in Subsection III.C, by the slow-switching control,
the population of the system
on the target state can reach at least
$\cos^2\frac{\theta}2$ .
Since
\begin{eqnarray}
\cos^2\theta\leq \cos^2\frac{\theta}2\leq 1, \ \ \  \theta \in[0,\frac{\pi}2],
\end{eqnarray}
we conclude that  all states after the slow-switching controls can be  steered to
the target by a single control. The free evolution time needed to accumulate the
relative phase is
\begin{eqnarray} \label{timeneeded}
t''=\frac{\phi''}{2\omega}=\frac 1{2\omega}\arctan
\frac{\cos E_+\tau'}{\sin E_+\tau'\cos\theta}, \label{time_needed}
\end{eqnarray}
where,
 \begin{eqnarray} \label{timeneeded2}
\tau'=\frac 1{E_+}{\arcsin\frac{\sin\frac{\gamma}2}{\sin\theta}},
\ \ \  \tau' \in [0,\frac{\pi}{2E_+}].
 \end{eqnarray}
When an initial state  $(\cos\frac{\gamma}2|e\rangle+
\sin\frac{\gamma}2|e\rangle)$ is very close  to the target state,
i.e., $\cos^2\frac{\gamma}2 \rightarrow 1$, by Eq. (\ref{timeneeded})
and Eq. (\ref{timeneeded2}), we   estimate that the relative
phase required for the single control is  $\phi'\rightarrow \frac
{\pi} 2$, corresponding to a free-evolution time $t''\rightarrow
\frac{\pi}{4\omega}$. This result is confirmed by
Fig.\ref{fig7}-(b), and it is in agreement with
the following relationship
\begin{eqnarray}\label{case1limit}
\frac{|b_{\tau}|^2}{|b_0|^2}\simeq \cos^2\phi'=\frac {1+\cos(2\phi')} 2,
\end{eqnarray}
since $|b_{\tau}|^2$  can be expanded in a
neighborhood of  $|b_0|^2$ when control starts with a state very close to its target, i.e.,
$b_0$ is very small.

In the case of small $b_0$, i.e.,
when $\phi' \rightarrow\frac{\pi}2$,
$$\frac{|b_{\tau}|^2}{|b_0|^2}\rightarrow 0, $$
with $\cos\frac{\gamma}2\rightarrow 1$.
This means if we let the system acquire a relative phase  $\phi'=\pi/2$,
the Lyapunov control would  become more  effective when the system is
very close to the  target state (see Fig.\ref{fig7}-(c)).


\section{conclusions}
We have studied the optimal Lyapunov control by an exactly solvable
two-level model in this paper. We found that the convergence time
and the control fidelity are  related to the strength of  control
fields. When the system is close to the target state, the
optimal Lyapunov control in (\ref{qubitcondition}) may become inefficient, i.e., the system under
control may not converge to the target state. To overcome this
difficulty, we extended the control law by combining free evolution
and external control. With this extended technique, the state in the vicinity
of target state can be controlled to the target by a single control.

\section*{Acknowledgments}
This work was supported by National Nature Science Foundation of China (NSFC) under grant
No.10905007 and No. 61078011, supported by the Fundamental Research Funds
for the Central Universities under grant No. DUT12LK28 and by the Australian Research Council (DP130101658, FL110100020).

\end{document}